\documentstyle[12pt,aasms]{article}

\tightenlines
\def\ni{\noindent}        

\def\hi{\noindent \hangindent=2.5em}
\def\et{{\it et\thinspace al.}}    
\def\pc{{\rm\,pc}}
\def\cm{{\rm\,cm}}
\def\kpc{{\rm\,kpc}}
\def\Mpc{{\rm\,Mpc}}

\def\Gyr{{\rm\,Gyr}}

\def\hnot{{\rm\,km/s/Mpc}}
\def\msun{{\rm\,M_\odot}}

\def\surfb{{\rm\,mag/arcsec^2}}
\def\surfden{{\rm\,M_\odot/pc^2}}
\def\numden{{\rm\,\#/Mpc^3}}
\def\araa{{\it Ann.\ Rev.\ Astr.\ Ap.}, }
\def\aj{{\it A.~J.}, }  
\def\apj{{\it Ap.~J.}, }  
\def\apjl{{\it Ap.~J.~(Letters)}, } 
\def\fcp{{\it Fund.~Cos.~Phys.}, } 
\def\mn{{\it M.N.R.A.S.}, }      
\def\nat{{\it Nature}, }      
\def\aa{{\it Astr.~Ap.}, }     

\def\spose#1{\hbox to 0pt{#1\hss}}
\def\lta{\mathrel{\spose{\lower 3pt\hbox{$\mathchar"218$}}
     \raise 2.0pt\hbox{$\mathchar"13C$}}}
\def\gta{\mathrel{\spose{\lower 3pt\hbox{$\mathchar"218$}}
     \raise 2.0pt\hbox{$\mathchar"13E$}}}

\def\clock{\count0=\time \divide\count0 by 60
     \count1=\count0 \multiply\count1 by -60 \advance\count1 by \time
     \number\count0:\ifnum\count1<10{0\number\count1}\else\number\count1\fi}

\begin{document}

\title{The Formation of Low Surface Brightness Galaxies}

\author{Julianne J. Dalcanton\altaffilmark{1,2}}
\affil{Princeton University Observatory,
       Princeton, NJ 08544 \\ \& \\
	Observatories of the Carnegie Institute
	of Washington, 813 Santa Barbara Street, Pasadena CA, 91101}

\author{David N.\ Spergel, \& F J Summers}
\affil{Princeton University Observatory,
       Princeton, NJ 08544}
\altaffiltext{1}{e-mail address: jd@ociw.edu}
\altaffiltext{2}{Hubble Fellow}

\begin{abstract}

The formation of low surface brightness galaxies is an unavoidable
prediction of any hierarchical clustering scenario.  In these models,
low surface brightness galaxies form at late times from small initial
overdensities, and make up most of the faint end of the galaxy
luminosity function.  Because there are tremendous observational
biases against finding low surface brightness galaxies, the
observed faint end of the galaxy luminosity function may easily fall
short of predictions, if hierarchical structure formation is correct.
We calculate the number density and mass density in collapsed objects
as a function of baryonic surface density and redshift, and show that
the mass in recently formed low surface brightness galaxies can be
comparable to the mass bound into ``normal'' high surface brightness
galaxies.  Because of their low gas surface densities, these galaxies
are easily ionized by the UV background and are not expected to appear
in HI surveys.  Low surface brightness galaxies (LSBs) are not a
special case of galaxy formation and are perhaps better viewed as a
continuance of the Hubble sequence.

\end{abstract}

\section{Introduction}

Over the past few decades, more and more machinery has been developed
to attempt to explain the distributions of galaxy luminosities and
morphologies that we see today.  A standard picture has developed in
which galaxies form through gravitational collapse of small density
perturbations.  In hierarchical models, the smallest astronomical
objects form early, with progressively larger objects forming at
progressively later times through merging of the earlier generations
of smaller objects.  Galaxies form fairly late in these models, with
clusters of galaxies assembling only in the most recent times.
Hundreds of analytical and numerical papers have explored the predictions
of hierarchical structure formation
and have been able to reproduce many of the properties of normal
galaxies and clusters (see for example White \& Frenk 1991, Efstathiou
\& Silk 1983, and the recent review by White 1994).
However, as discussed in these papers, hierarchical structure
formation consistently predicts many more faint galaxies than are
actually observed.  The faint-end of the predicted luminosity function
is much steeper than luminosity functions generated from catalogs of
nearby galaxies.

Simultaneously, there has been increasing attention focused on the
existence of an often overlooked population of low surface brightness
galaxies (LSB's).  Because of the brightness of the night sky,
observers are naturally biased towards detecting high-surface
brightness galaxies, whose high contrast against the background makes
them easily detectable.  This strong selection effect, noted by Zwicky
(1957) and further explored by Disney (1976), can lead to strong
correlations in the properties of observed galaxies that are not
intrinsic to the objects (Disney 1976, Allen \& Shu 1979), most
notably the universal surface brightness of spiral disks discovered by
Freeman (1970).  Whenever the bias against finding low surface
brightness galaxies has been reduced, however, LSB's have appeared
which violate the Freeman relation (see surveys by Impey, Bothun,
\& Malin 1988, Bothun, Impey,
\& Malin 1991, Schombert et al. 1992, Schombert \& Bothun 1988,Irwin,
Davies, Disney, \& Phillipps 1990, Turner, Phillipps, Davies, \&
Disney 1993,  Sprayberry et al 1995, and references therein).  These surveys
show that LSB's do exist and have been  previously overlooked as
a potentially significant species in the galaxy menagerie.
The existence of elusive
but omnipresent LSB's implies that we have been attempting to solve the
puzzle of galaxy formation with many pieces missing.

Thankfully, our ignorance is succumbing to a growing body of
observations of low surface brightness galaxies.  First, LSB's exist
at every surface brightness to which surveys have been sensitive.
They have been detected down to central surface brightnesses of 26.5
mag/arcsec$^2$ in $V$ (Dalcanton 1995); fainter than this, it is
difficult to separate LSB's from true fluctuations in the optical
extragalactic background due to distant clusters of galaxies (Shectman
1973, Dalcanton 1994).  Second, LSB's exist at every size, from minute
dwarf galaxies in the Local Group, through galaxies with scale lengths
typical of ``normal'' high surface brightness (HSB) galaxies (McGaugh
\& Bothun 1994), up to a handful of truly giant galaxies with scale
lengths of $\approx 50$ kpc, typified by Malin I\footnote{The
exceptional giant ``Malin-type'' galaxies, while being fascinatingly
odd, are unlikely to be the dominant type of LSB galaxy.  Their large
sizes imply that 2-dimensional surveys have extremely large accessible
volumes for this particular type of object.  The fact that so few have
been found (for example, see Figure 3 of Schombert et al.  (1992))
immediately suggests that their number density is small.  If these
objects are the result of large fluctuations in low density regions
(as postulated by Hoffman, et al 1992), then they ought to be rare.}
(Bothun et al. 1987).  Third, LSB's are roughly a factor of two less
correlated than HSB galaxies in the CfA and IRAS catalogs on scales
$>1\Mpc$ (Mo, McGaugh, \& Bothun 1994), and even less correlated on
smaller scales (Bothun et al. 1993).  Finally, extensive studies of
LSB colors (McGaugh \& Bothun 1994, Knezek 1993) and spectroscopy of
HII regions (McGaugh 1994, McGaugh \& Bothun 1994) show that LSB's
have rather blue colors (although with large scatter) that cannot be
explained by either low metallicity or high current star formation
rates; the blue colors are more readily explained by LSB's having a
relatively young mean age with a relatively long time-scale for star
formation.

The properties of known LSB's fit naturally with their likely
formation scenario.  Their clustering properties and long formation
timescales immediately suggest that LSB's form from the collapse of
smaller amplitude overdensities than normal HSB galaxies (Mo, McGaugh,
\& Bothun 1994), with the exception of Malin-type LSB's which likely
form from rare, isolated $3\sigma$ peaks (Hoffman, Silk, \& Wyse
1992).  For a simple top-hat collapse of an overdense region, small
amplitude peaks in the background density take longer to reach their
maximum size and longer to recollapse than higher amplitude peaks,
implying that galaxies that collapse from the smaller peaks will have
later formation times and longer collapse times.  In any
theory with Gaussian fluctuations, small amplitude galaxy-sized peaks
are more likely to be found in underdense regions, where their mean
level is pulled down by the large scale underdensity, suggesting that
objects which collapse from small amplitude peaks will be less
correlated than those that collapse from larger ones (Kaiser 1984,
White et al. 1987).  These are exactly the properties being uncovered
for LSB galaxies.

In this paper, we expand upon this idea, showing that LSB's are a
general {\it prediction} of existing hierarchical
theories of structure formation.  LSB's naturally make up most of the
faint end of the predicted galaxy luminosity function, which, given
the severe underrepresentation of these galaxies in all existing
catalogs, explains how current low measurements of the faint end of
the luminosity function can be reconciled with the theoretical
predictions of large numbers of low luminosity galaxies.  Furthermore,
we show that the mass density in recently forming LSB's can be
comparable to the mass in ``normal'' galaxies, particularly in models
of high bias.  In \S\ref{otherastro}, we conclude
with a discussion the astrophysical implications of the existence of
large population of LSBs, in particular considering HI surveys, Lyman-$\alpha$
absorbers, the faint blue galaxy excess, and the Tully-Fisher
relation.

\section{The Formation of LSB's}

\subsection{Theoretical Assumptions}

In the standard gravitational instability picture for galaxy
formation, initial overdensities in the distribution of mass expand
more slowly than the universe as a whole, eventually separating from the global
expansion and collapsing onto
themselves (Lifshitz 1946).  The collapsed
regions still continue to grow roughly isothermally in size and mass,
as successively larger shells of material themselves collapse onto
the initial overdensity and virialize, or as adjacent collapsed
regions merge together.  It is straight-forward to track the collapse of the
non-dissipative material to form a dark halo.
The collapse of the baryonic matter,
however, is more complicated.  The baryons are dissipative, subject
to pressure, heating, cooling, and feedback due to star formation.
They undergo a more complicated collapse within the dark matter
halo to form the stellar disks and ellipsoids.

Because the surface brightness of a galaxy will be more closely linked
to the baryonic surface density than to the dark matter surface
density, we must find a way to relate the easily calculated collapse
of the dark matter to the collapse of the baryons.  Thankfully, with a
few reasonable assumptions, the structure of the dissipative baryonic
matter can be simply related to the structure of the dark halo (Faber
1982). These assumptions are (i) the baryonic fraction within initial
overdensities is constant and (ii) the net angular momentum that the
baryonic component acquires during the collapse is ultimately
responsible for halting the collapse.  The first assumption is
supported by the hydrodynamic simulations of Evrard, Summers, \& Davis
(1994), which finds a roughly constant ratio between the baryonic and dark
matter for all galaxies (see their Figure 12).  The second assumption
appears to be valid for disk systems that are supported by rotation
rather than by random motions.  Although we will use the term
``baryonic'' interchangeably with ``dissipative'' for the duration of
this paper, we recognize that there may be baryons which have been
bound into an early generation of stars (Population
III) or black holes and thus will not participate in a dissipative collapse.

With these assumptions, if the non-baryonic component of a shell
collapses to form an isothermal halo of radius $r_H$, then the
baryonic component collapses to a radius of $r_*$ where
the collapse factor $\Lambda(\lambda)$ is defined to be

\begin{equation}				\label{collapse}
	\Lambda^{-1}(\lambda) \equiv {r_* \over r_H}
			= \lambda \, [\sqrt{(F/2\lambda)^2 + 2}
				     - (F/2\lambda)]
\end{equation}

\ni (Faber 1982). $F$ is the ratio of the dissipative baryonic
mass to the total mass within the
shell, and $\lambda$ is the dimensionless spin parameter

	   \begin{equation}
		\lambda = {\cal L} \, |E|^{1/2} \, G^{-1} \, M^{-5/2}
	   \end{equation}

\ni (Peebles 1969).
Here, ${\cal L}$ is total angular momentum of the luminous and
non-luminous matter, $E$ is its energy, and $M$ its mass.  Because the
baryonic component collapses to a fixed fraction of the size of the
non-baryonic component (for a given value of $\lambda$), we may now
use the properties of the dark matter halos to predict the properties
of the visible matter.

If we further assume that galaxies, on average, have a similar ratio
between their baryonic mass and their luminosity (i.e., a similar
efficiency in converting baryons to light-producing stars), then we
now have a way to relate the surface density of the dark halo to the
surface brightness of the luminous galaxy.  While other scenarios may
argue that the low surface brightness of LSB's is a result of baryonic
physics that we do not consider in this calculation (for example, low
star-formation efficiencies due to low metallicity or the absence of
tidal triggering, gas loss through supernovae explosions or ionization
by the UV background), we prefer to explore the straight forward assumption
that a galaxy with fewer baryons per unit area
will in general form fewer stars per unit area. For the purposes of
this paper, we relegate this rich astrophysics to being only a
perturbation to a global proportionality between baryonic surface
density and surface brightness.  The results will show
that even with this naive assumption, the bulk properties of LSBs can
be understood as the inevitable result of the properties of low
surface density galaxies.

We would venture that our neglect of detailed physical processes
within the galaxies causes us to systematically overestimate the
surface brightness of LSBs; almost all mechanisms for reducing the
star formation efficiency are likely to be more effective in low
surface density galaxies than in high surface density ones.  First, if
star formation requires a large reservoir of neutral gas, then the
extragalactic ultraviolet background, which will completely ionize low
surface density galaxies (\S\ref{HI}), may suppress or shut off the
formation of stars in these galaxies (Babul \& Rees 1992).  Second, if
star formation is associated with disks becoming Toomre unstable to
the formation of spiral structure (implying that the Toomre stability
parameter $Q\propto \sigma\kappa / G\Sigma$ is smaller than some
constant, where $\sigma$ is the velocity dispersion of the gas,
$\Sigma$ is the surface density of the disk, and $\kappa$ is the
epicyclic frequency, set by the disk structural parameters (see
Kennicutt 1989)), then one would expect lower surface density disks to
be less unstable for star formation.  Observations by van der Hulst et
al. (1993) find that LSB galaxies do have HI surface densities that
that fall below the critical density implied by $Q$ and are about a
factor of 2 lower than the HI surface densities of HSB galaxies.
Third, if star formation is enhanced by tidal interactions with nearby
galaxies, then galaxies that have fewer close neighbors will have
lower star formation efficiencies.  We will argue that, as seen
observationally and discussed by (Mo, McGaugh, \& Bothun 1994), low surface
density
galaxies should have lower correlation amplitudes and thus fewer close
neighbors, which would once again lead to lower surface
brightnesses\footnote{Note that the case for causality is a bit
ambiguous.  The impression that close neighbors induce higher star
formation rates could be an artifact of low surface density galaxies
being less correlated than high surface density ones, with no
dependence of the star formation efficiency on environment.}.  Fourth,
low surface density, low mass galaxies are more likely to lose their
gas through supernova driven winds than high surface density, high
mass galaxies.  This both shuts off star formation prematurely and
evolves the galaxy towards larger sizes through sudden mass loss and
subsequent revirialization (Dekel \& Silk 1986, DeYoung
\& Heckman 1994) -- both mechanisms which lead to lower surface
brightnesses for low surface density objects.

In spite of our previous assertion to the contrary, there is one piece
of baryonic physics which we cannot ignore, namely pressure.  We are
ultimately interested in the gravitational collapse of baryons, but a
collapsing baryonic gas both the inward tug of
gravity and a resistive force due
to its own internal pressure.  For sufficiently large masses, the
self-gravity of the combination of baryons and dark matter is strong
enough for the effects of pressure to be negligible.  However, for
small masses the pressure is sufficient to support the baryonic gas
against collapse, in spite of the additional gravitational pull
provided by the collapsed dark matter halo.  Because low surface
density galaxies are more likely to have lower total masses than high
surface density galaxies, we must consider the mass scales on which
the effects of pressure become important; we will do so in
\S\ref{jeans}.

\subsection{Properties of LSBs}

We now examine the conditions that lead to variations in
baryonic surface density (and thus surface brightness)
among galaxies.
We choose to compare the properties of galaxies within a  fixed radius
rather than at a constant mass scale.

At constant mass, comparing LSB's to HSB's is a
comparison between galaxies of widely different scale lengths: high
surface brightness dwarf ellipticals would be compared to giant
low-surface brightness galaxies.  By comparing the surface brightness
of galaxies at constant mass, one is saying that what distinguishes
LSB's from HSB's is that LSB's have abnormally large scale lengths for
their mass.  Instead, we will compare galaxies of different
masses, but with similar scale lengths.  This more closely mimics how
one draws the distinction between LSB's and HSB's.
We therefore define the effective surface density, ${\bar\Sigma}$ to be the
mean surface
density within some radius $r_*$,

\begin{equation}				\label{sigma_m}
{\bar\Sigma} = { M(r<r_*) \over {\pi\,r_*^2}}.
\end{equation}

\ni where $M(r<r_*)$ is the mass of the baryonic
component within a radius $r_*$.  We can relate $M(r<r_*)$ to the
total mass of the halo,

\begin{equation}				\label{mass}
M(r<r_*) = F \, M_{tot}(r<r_H) = F \, {{4\pi}\over 3} \, \rho_0 \, r_0^3,
\end{equation}

\ni where $\rho_0$ is the current density of the universe and assuming
that the dark matter that initially was in a shell of comoving radius
$r_0$ has collapsed to a virial radius $r_H$, while baryons from the
same shell have collapsed and dissipated to a radius $r_*$, where they
are supported by their angular momentum.  Equations~\ref{sigma_m} and
{}~\ref{mass} imply that $r_0$ is determined by the choice of
${\bar\Sigma}$ and $r_*$:

\begin{equation}				\label{r0_sigma}
 r_0({\bar\Sigma},r_*) = \, r_* \,
			\left( {3 {\bar\Sigma} \over 4 {F\,\rho_0\,r_*}}\right)^{1/3}
\end{equation}

In turn, the redshift of collapse can be determined
from $r_0$ and $r_*$ as follows.  In a spherical symmetric top-hat collapse
(Gott and Gunn 1972, Peebles 1980), the dark matter virializes at a
radius of half its size at maximum expansion, the density within a shell at
maximum
expansion is $(3\pi/4)^2$ times the background density, and  the
collapse time is twice the time of maximum expansion.
Therefore, for top-hat collapse in a $\Omega=1$ universe,

\begin{eqnarray}
	r_0 &=& r_H \, (18\pi^2)^{1/3} \, (1 + z_c)  \\
	    &=& r_* \, \Lambda \, (18\pi^2)^{1/3} \, (1 + z_c)  \label{r0_rH}
\end{eqnarray}

\ni where $z_c$ is the redshift at which the shell has collapsed
and virialized.  Equations \ref{mass} and \ref{r0_rH} may be
substituted into equation~\ref{sigma_m} to solve for the redshift of
collapse of a galaxy with mean baryonic surface density ${\bar \Sigma}$ within
$r_*$:

\begin{equation}				\label{zc}
	z_c({\bar\Sigma},r_*) =
			\left({{\bar\Sigma} \over {24 \pi^2
		F\,\rho_0\,r_*}}\right)^{1/3} \,
	  		\Lambda^{-1}(\lambda)	 -  1.
\end{equation}

Alternatively, this equation may be rearranged to express ${\bar\Sigma}$ in
terms
of $z_c$:

\begin{equation}		\label{sigma_z}
		{\bar\Sigma}=24\pi^2 \, F \, \rho_0 r_* \, \Lambda^3(\lambda) \,
			(1+z_c)^3,
\end{equation}

\ni which immediately implies that low surface density galaxies form
at later times than high surface density galaxies with the same
angular momentum.  If, as we have asserted, surface density is
proportional to surface brightness, then low-surface brightness
galaxies have formed more recently than high-surface brightness
galaxies of similar size.  This also suggests that for galaxies of a
given size and angular momentum, there is a minimum surface density
${\bar\Sigma}_0=24\pi^2\,F\,\rho_0\,r_* \,\Lambda^3(\lambda)$ below which
few galaxies should exist.

To derive equation~\ref{sigma_z}, we have ignored infall due to the
collapse of shells larger than $r_0$ after $z_c$.  This is not a bad
assumption for the dark matter; the larger shells virialize at larger
radii, and thus the increase in mass occurs primarily at large radii.
Baryons are dissipative, however, and can collapse until halted by
their angular momentum.  Including the effects of infall at late times
would only exacerbate the surface brightness distinction between
early-forming HSB's and late-forming LSB's, as HSB's would have more
time to accrete mass and further increase their surface brightness.
Thus, late-time infall likely enhances rather than erases the
correlation between formation time and surface brightness.
Simulations of Evrard et al.\ (1994) justify our neglect of late time
infall as they find that the rate at which galaxies accrete mass slows
dramatically at late times.

It is revealing to note that we have implicitly made the assumption
that there is a linear correspondence between a galaxy's mean surface
brightness and its luminosity (eq.~\ref{sigma_m}, assuming a single
mass-to-light ratio).  This is an artifact from associating a fixed
physical scale $r_*$ with all galaxies; for example, if instead one
were to consider the luminosity within some isophotal level, the
linearity between surface brightness and luminosity would break down.
There are also additional sources of scatter in the relationship which
we have ignored: variations in angular momentum, variations in
collapse time due to changes in the mean local overdensity, variations
in the mass-to-light ratio with mass, and deviations from spherical
collapse.  Regardless of these complications, it is difficult
to conceive of eq.~\ref{sigma_m} not being a reasonable first-order
approximation to the relationship between surface density
(i.e. brightness) and mass (i.e. luminosity); certainly the trend is
displayed when considering spirals and ellipticals.
The relationship between surface brightness and luminosity has immediate
implications for redshift
surveys.  If a redshift survey has a limiting surface brightness for
spectroscopy, set either by design or by the limits of the
spectrograph, then there will be an associated limiting {\it
luminosity} to the galaxies that are observed.  Decreasing the
magnitude limit of a spectroscopic survey without decreasing the
surface brightness limit will lead one to pick up galaxies of the same
luminosity as were observed previously, only further away.  This
will inevitably lead to underestimates in the derived faint-end slope
of the luminosity function, as well as variations among different
surveys which have different, often unstated, surface brightness limits.

\subsection{Angular Momentum and Profiles of LSBs}	\label{profiles}

In addition to suggesting that galaxies that form late are more likely
to be LSB's, Equation~\ref{sigma_z} also implies that LSBs could be
galaxies whose baryons have not collapsed much further beyond the
non-baryonic dark halo.  Galaxies whose baryons have collapsed very
little must have either acquired large amounts of angular momentum
during their formation, or have been inefficient at transporting
angular momentum outwards during their dissipative collapse.  Both these
traits can be associated with galaxies that form at
lower redshifts.

First, galaxies that formed late are unlikely to have extremely
low angular momenta.  Analytic calculations and numerical simulations
have shown that galaxies acquire angular momentum through tidal torquing.
While systems that form very early, such as globular
clusters and (presumably) most ellipticals, may have collapsed so quickly that
they had little
time to acquire angular momentum, late forming galaxies have had
ample time to be torqued by nearby galaxies and by the shear in the
global gravitational field.  There have been suggestions of this trend
in numerical work, but it has not yet been well quantified, particularly
for the low amplitude peaks that are associated with LSBs.  Eisenstein
\& Loeb (1994, private communication) find that, on a given mass scale,
the value of $\lambda$ is anti-correlated with peak height, leading
to a correlation between angular momentum and collapse times (see eq.
\ref{sigma_delta}).  They
suggest that the relation arises because the separate axes of
an object with a small overdensity collapse at very different times, giving
a larger quadrupole moment and making the object more susceptible
to external torques.

Second, the trend from ellipticals to spirals suggest that early
forming galaxies have been particularly efficient at shedding angular
momentum.  While ellipticals appear to have collapsed dramatically,
they have very low angular momenta, suggesting that large amounts were
transported outwards during collapse.  Spiral disks, which are in
general younger systems than ellipticals, have large angular momenta,
suggesting that the dramatic angular momentum transport that was
operating during the epoch of elliptical formation was much less
spectacular during the formation of spiral disks.  This trend between
angular momentum transport and formation epoch may have several
different origins.  It could reflect changes in the galaxy environment
with time (e.g.  higher external pressure, more gravitational shear,
evolution in the triaxiality of halos) or it could reflect processes
within the galaxy halo that depend upon the duration of collapse time
(e.g. feedback from star formation, interactions between the halo and
baryons).  However, regardless of the origins of the trend, it can be
extrapolated to the present day to suggest that any galaxies that are
currently forming are more likely to have inefficient angular momentum
transport.  If this were not true, then one might expect a large population of
very young, blue elliptical galaxies to be forming today.  Except for the
very smallest dwarf galaxies ($M_B > -14$), the young blue galaxies
that are observed tend to be irregular galaxies and be supported by
rotation (Lo, Sargent, \& Freeman 1993, Gallagher \& Hunter 1984, and
references therein).  We note that classical ellipticals are exempted from much
of the discussion in this paper.  Because they have obviously not
conserved angular momentum during their collapse, they do not
obey equation \ref{collapse} or any equation that follows from
it.

When angular momentum transport is inefficient, the distribution of
angular momentum per unit mass is likely to be constant during
collapse, which leads to the formation of exponential disks (Gunn
1982).  Thus, if LSBs formed late, as suggested by
equation~\ref{sigma_z} and observed stellar populations (Knezek 1993,
McGaugh 1994, McGaugh \& Bothun 1994), then they should be disk
systems with exponential profiles.  For non-dwarf LSBs, exponential
profiles are an excellent fit to the stellar distribution in both
infrared and visible bands (Davies, Phillips, \& Disney 1990, McGaugh
\& Bothun 1994, James 1994, McGaugh et al 1995, Dalcanton 1995).
Alternatively, this argument can be inverted and the ubiquity of
exponential disks in LSBs can be used to argue for inefficient angular
momentum transport and large values of $\lambda$.

With this view of LSB's, one can interpret LSB's as being an extension
of the Hubble sequence.  At one end of the sequence are ellipticals,
consisting entirely of highly collapsed, low angular momentum systems
with short formation timescales.  In the middle of the sequence are
spiral galaxies, with an increasing fraction of high angular momentum
disk stars, an increasing timescale for star formation, and lower
masses.  LSB's could be the obvious next step in the sequence, with
even longer star formation times, higher angular momenta, and smaller
masses and bulges.  In the absence of strong arguments for why disks
should have stopped forming past the Sc end of the Hubble sequence,
and in the undeniable presence of rapidly rotating low surface
brightness disks (Schombert et al 1992, Sprayberry et al 1995), it is
difficult to argue that the Hubble sequence truly ends at its standard
terminus.

\subsection{Correlations of LSBs}		\label{correlations}

If LSBs formed at late times from low amplitude fluctuations, then the
LSBs should be less correlated than HSBs, which formed from earlier
from higher amplitude fluctuations.  Thus, LSBs extend the
morphology-density relationship seen within elliptical and spirals:
high overdensities implies early formation and high correlation, which
in turn implies high surface brightness. Recasting
equation~\ref{sigma_z} in terms of $\delta = \delta\rho/\rho$, the
initial overdensity from which the galaxy collapsed,

\begin{equation} 	  			\label{sigma_delta}
	   {\bar\Sigma} \propto F\, r_* \, \Lambda^3(\lambda) \,\delta^3.
\end{equation}

\ni Here, we assume for simplicity, $\Omega = 1$ and thus,
$\delta \propto (1+z)^{-1}$.  In a Gaussian theory, the probability
of finding two peaks of amplitudes between $\nu\sigma$
and $(\nu + \epsilon)\sigma$ within a sphere of radius of $r$ is:

\begin{equation}
P_2 = {1 \over 2 \pi} {\epsilon^2 \sigma^2 \over
\sqrt{\xi(0)^2 - \xi(r)^2}} \exp\left[-{\nu^2 \sigma^2 \over
\xi(0)+ \xi(r)}\right],
\end{equation}

\ni where $\epsilon << \nu$ and $\sigma^2$ is the variance of the
Gaussian fluctuations (Kaiser 1984).
Recall $\xi(0) = \sigma^2$ and that this assumption of Gaussian
fluctuations ignores the non-linear evolution of the density field.
As the probability of finding a single peak within
the sphere, $P_1$ is $(\epsilon/\sqrt{2\pi}\sigma) \exp(-\nu^2/2)$,
the correlation of peaks of amplitude $\nu$ is

\begin{equation}
\xi_\nu(r) = {P_2 \over P_1^2} -1 = {\xi(0)^2
\over \sqrt{\xi(0)^2 - \xi(r)^2}} \exp\left[{\nu^2 \xi(r) \over \xi(0) + \xi(r)
}\right] - 1
\end{equation}

\ni At large separation, where
$\xi(r) << \xi(0)$, then

\begin{equation}		\label{correlation}
	\xi_\nu(r) = {\nu^2 \over \sigma^2} \xi(r)
\end{equation}

\ni Thus, the proportionality between fluctuation amplitude and
correlation strength is expected not only for the high amplitude peaks
that form clusters (Kaiser 1984), but for all collapsed objects.
Equations~\ref{correlation} and~\ref{sigma_delta} predict that LSBs
are less correlated than HSBs, consistent with the Mo, McGaugh, \&
Bothun (1994) analysis of observed LSB and HSB samples.  Further
evidence of this reduced correlation may be manifested in weak
correlation of faint galaxies (Koo \& Szalay 1984, Stevenson et al.\ 1985,
Efstathiou et
al.\ 1991, Pritchett \& Infante 1992, Bernstein et al.\ 1993), given
that the deep observations are more sensitive to LSBs than local
surveys (Phillips, Davies, \& Disney 1990, McGaugh 1994, Ferguson \&
McGaugh 1995).

The relationship between ${\bar\Sigma}$ and $\delta$ given in
equation~\ref{sigma_delta} also explains why differences in the
correlation function of LSB's and HSB's have only recently become
apparent with the development of new samples of truly low-surface
brightness galaxies.  Because surface density is a strong power
of the initial overdensity, only a sample with a large range of
surface densities would manifest properties that depend on a weaker
power of $\delta$.  The amplitude of the correlation function traced
by galaxies is proportional to $\delta^2$ (eq.~\ref{correlation}), and
thus proportional to ${\bar\Sigma}^{2/3}/\Lambda^{2}(\lambda)$.  This
suggests that to detect a 50\% difference between the correlation
amplitudes for HSB's and for LSB's, as was seen in Mo, McGaugh, \&
Bothun (1994), one needs the LSB sample to be cleanly separated in
surface brightness from the HSB sample by at least one magnitude per
square arcsecond, and possibly more if low surface brightness galaxies
have smaller values of the collapse factor $\Lambda(\lambda)$.  The required
range of
surface brightness did not exist in earlier work comparing the CfA
galaxies to galaxies drawn from the UGC catalog (Thuan, Alimi, \& Gott
1991), nor in the work of Bothun et al. (1986) where the larger range
in surface brightness was swamped by large uncertainties in the
measurement of the surface brightnesses.  Only by pressing surveys for
LSB's to the lowest possible surface brightnesses would one hope to
uncover a population of galaxies encroaching upon the voids.

\section{The Surface Brightness Distribution of Galaxies}  \label{surfbriden}

With the formulas developed above, we are now in a position to
calculate the number density of galaxies as a function of surface
density and redshift.  We will use the Press-Schechter formalism,
which uses linear theory to calculate the number density of regions of
initial radius $r_0$ whose extrapolated linear overdensities are
sufficiently large that the region has in fact undergone a non-linear
collapse to form a virialized object at redshift $z_c$ (Press \&
Schechter 1974).  The formalism assumes that the initial fluctuations
are Gaussian and that the overdense regions undergo a spherical
top-hat collapse which stops when the systems virialize at a
radius of half the radius at maximum expansion (see Peacock \& Heavens
1989, Bond et al.\ 1991, \& Bower et al 1991 for
discussions of this formalism).  With these assumptions,
the number density of objects with initial radius $r_0$ that have just
collapsed at $z_c$ is,

\begin{equation}				\label{press-schechter}
	n_{PS}(r_0,z_c) = -\left({2\over \pi}\right)^{1/2} \,
		\left[{{\delta_c(1+z_c)}\over{\Delta^2(r_0)}} \,
			  {{\partial\Delta(r_0)}\over{\partial r_0}} \,
			     e^{-{ {\delta_c^2 (1+z_c)^2} \over
				   {2\Delta^2(r_0)}}} \right]
		\, \frac{\rho_0}{M(r_0)},
\end{equation}

\ni where $M(r_0)$ is the mass within an initial radius $r_0$ and
$\Delta(r_0)$ is the variance in density within shells of radius
$r_0$ for a power spectrum of fluctuations $P(k)$, defined as

\begin{equation}
	\Delta^2(r_0) = \int^\infty_0 4\pi k^2 \, dk P(k) W^2(kr_0)
\end{equation}

\ni where $W(x) = 3(\sin{x} - x\cos{x}) / x^3$
and $\delta_c$ is the extrapolated
linear overdensity that the clump would have had at $z_c$ if
it had not collapsed, taken to be 1.68 to agree with numerical simulations.
For a CDM power-spectrum, we use

\begin{equation}
	P(k) = 1.94\times10^4 \, b^{-2} \,
			k(1+6.8k + 72k^{3/2} + 16k^2)^{-2} \Mpc^3,
\end{equation}

\ni for $H_0=50$ km/s/Mpc, taken from Davis et al. (1985).  The bias
parameter, $b$ is defined to be the ratio between the variances of the
galaxy and the mass fluctuations within 16 Mpc radius spheres.  This
particular approximation is accurate to 10\% for scales between
$0.05\Mpc$ and $40\Mpc$, and is too high on small scales.

The collapsed objects traced by equation~\ref{press-schechter} never
stop increasing their mass; they continue to accrete matter from
progressively larger shells which continue to collapse around and
merge into the object.  Therefore, something that we might identify as
a single object is associated with a different mass and different
$r_0$, depending upon the redshift at which we choose to identify it.
If we associate these ``collapsed objects'' with the more prosaic term
``galaxy'', we immediately see how difficult it is to define exactly
when a galaxy has formed.  In the absence of any physics beyond
gravity, there is no particular scale that naturally selects the
criteria for labelling a galaxy.

As motivated by our discussion of surface brightness above, we will
chose a radius criterion for deciding when a galaxy forms; we will say
that a galaxy has formed when the baryons originating in some
particular shell collapse to a final size $r_*$.  For a given surface
density at the time of formation, the choice of $r_*$ fully specifies
both the redshift at which the galaxy formed (eq.~\ref{zc}), and the
initial size of the shell $r_0$ from which the galaxy formed (i.e. the
shell whose baryons collapsed to size $r_*$ at $z$)
(eq.~\ref{r0_sigma}).  Equation~\ref{r0_rH} can be used with equation
\ref{press-schechter} to calculate the number density of galaxies
that have already collapsed by redshift $z$ from an initial radius
$r_0$ to a final radius $r_*$, with spin parameter $\lambda$:

\begin{equation}
n_r(r_0,\lambda|z) = n_{PS}(r_0,z_c(r_0,\lambda))
			\, p(\lambda|z_c(r_0,\lambda))
			\, \left|\frac{\partial z_c}{\partial r_0}\right|
			\, \Theta(z_c(r_0,\lambda)-z),
\end{equation}

\ni where $p(\lambda|z)$ is the probability that a galaxy forming at
redshift $z$ has a spin parameter $\lambda$, and where $\Theta(x)=1$
if $x>0$, and equals zero otherwise.  For the sake of notational
simplicity, we do not make the dependence on $r_*$ explicit.

Transferring variables from $r_0$ to ${\bar\Sigma}$ (eq.~\ref{r0_sigma}), and
integrating over $\lambda$ gives the total number density of galaxies
that have formed by $z$ with a given surface density ${\bar\Sigma}$:

\begin{equation}					\label{n_sigma_z}
	n({\bar\Sigma}|z) = \int^\infty_0
		n_r\left(r_0({\bar\Sigma}),
			 \lambda |
			 z<z_c\left(r_0({\bar\Sigma}),\lambda\right)\right)
			\left|{ {\partial r_0} \over {\partial {\bar\Sigma}}}\right|
			\, d\lambda,
\end{equation}

\ni where $z_c\left(r_0({\bar\Sigma}),\lambda\right)$ reduces
to equation~\ref{zc}.

Analytical calculations and numerical simulations find that large
protogalaxies typically form with values of $\lambda\sim0.05\pm0.05$
(Peebles 1969, Barnes \& Efstathiou 1987, Warren et al.\ 1992,
Steinmetz \& Bartelmann 1994, Eisenstein \& Loeb 1994).  Instead of
choosing one particular, highly uncertain model for $p(\lambda|z)$,
we find it more illustrative to assume single fixed values for
$\lambda$:$p(\lambda|z)=\delta(\lambda-\lambda_0)$.
In \S\ref{profiles} we argued that LSBs can be treated as late-forming,
high angular momentum extensions to the Hubble sequence, and as such
we should consider values of $\lambda$ that are appropriate for the
end of the Hubble sequence and beyond.  Following an argument given by
Faber (1982), the measured fraction of baryonic mass within the
optical radius ($\equiv X$), can be used to estimate the factor by
which the baryons have collapsed.  For collapse within a fixed,
non-dissipative, isothermal halo, the baryonic collapse factor is

\begin{equation}
\Lambda = \left[\frac{X}{1-X}\right] \, \left[\frac{1-F}{F}\right]
\end{equation}

\ni For the values of $X$ given in Faber's Table 3 for Sa, Sc, and
Irr galaxies\footnote{drawn from Faber \& Gallagher (1979), Thuan \&
Seitzer (1979), and Roberts (1969)}, the collapse factors are roughly
100 for the Sa's, 10-15 for the Sc's, and upper limits of 7-11 for the
Irr's, which corresponds to $\lambda=0.02,\,0.06-0.09,\,>(0.09-0.12)$,
assuming $F=0.05$.  Based on this, we examine cases where
$\lambda=0.075,0.15$, which are the 90\% and 99\% percentiles of the
distribution of $\lambda$ found numerically by Eisenstein \& Loeb
(1994) for $>2.5\sigma$ peaks with $M=10^{12}\msun$.  Note that
types Sc and later make up 30\% of the number density of galaxies with
$L/L_*>0.1$ (Marzke et al 1994), a larger fraction than would be
implied by the distribution of $\lambda$ in Eisenstein \& Loeb (1994).
This suggests that numerical simulations may either underestimate the
fraction of high-spin halos or that baryons acquire more specific
angular momentum during their collapse than does the dark matter.  It
could also be taken as evidence that galaxies with types later than
Sc form smaller overdensities than were assumed in Eisenstein \&
Loeb (1994), and thus form in greater numbers and with larger
values of $\lambda$.

The resulting distributions are plotted in the first two columns of
Figures 1 \& 2 for $b=1$ \& $2.5$, $z=0-5$ (light to dark) with
$\lambda=0.03,0.075,0.15$ from
the top row to the bottom.  To interpret the distributions, it is
necessary to first establish some tie with ``normal'' galaxies (by
which we mean high surface brightness, nearby, cataloged galaxies with
$L/L_*>0.1$) through their redshift of formation, angular momenta,
surface densities, and number densities.  First, we choose to
associate normal galaxies with formation times of $z=2$ or earlier.
The lookback time to $z=2$ is $7-8h_{75}^{-1}\Gyr$ for $\Omega=0.2-1$
and $h_{75}=H_0/75\hnot$.  In the Milky Way, the ages of F and G
dwarfs show that a significant population of stars in the disk were
formed over $8\Gyr$ ago from high metallicity gas, suggesting that the
disk of the galaxy had already settled into place to some degree by
$z=2$ (Barry 1988, Carlberg et al.\ 1985, Twarog 1980), bolstering our
identification of this epoch as being the one by which the mass of
galaxy disks had been assembled.  Second, we chose $\lambda=0.075$ to
be the fiducial case for identifying ``normal'' galaxies because it
roughly demarks the maximum spin angular momentum of the most numerous
classes of galaxies (Sd and earlier).  Galaxies that form with smaller
values of $\lambda$ will all form with larger surface densities, and
thus a galaxy that forms with $\lambda=0.075$ can be used to define the
limit where normal surface brightnesses end, and low surface
brightnesses begin.  Furthermore, by choosing a value of $\lambda$
that corresponds to galaxies with very small spheroidal components, we
hopefully avoid the need to determine the fraction of baryons that
wind up in the disk rather than the bulge.

We use the Milky Way to estimate the surface density associated with
normal galaxies.  If the baryonic surface density in the solar
neighborhood is $75\surfden$, the mean baryonic surface density of
the Milky Way is roughly $200-300\surfden$ for $r_*=7.5-10\kpc$,
including a 20\% correction for the mass in the bulge.  (For a disk
mass-to-light ratio of 5, this gives the correct Freeman disk central
surface brightness.)  In Figures 1 \& 2 we have chosen $r_*$ such that
for $\lambda=0.075$ the integrated number density of galaxies that
have collapsed by $z=2$ is roughly the total number density of normal
galaxies measured today (corresponding to the horizontal dashed and
dotted lines in the second columns of Figures 1\& 2.  Note that this
choice of $r_*$ automatically gives surface densities for the galaxies
forming at $z=2$ that are in good agreement with the Milky Way value.
Galaxies that form before $z=2$ with larger surface densities are only
a small fraction of the number density at $z=2$; 75-99\% of the
galaxies that form by $z=2$ (for $b=1-2.5$) have collapsed between
$z=2$ and $z=3$.

The particular form of $n({\bar\Sigma}|z)$ seen in Figures 1 \& 2
arises because, one, we've assumed that the surface density of a
galaxy doesn't change after it forms, and two, galaxies of a given
surface density all form at the same redshift.  With these two
assumptions, the only change in the distribution of surface densities
with redshift is the creation of new galaxies at increasingly lower
surface brightnesses (eq.~\ref{sigma_z}).  Because low surface density
galaxies form from low amplitude peaks, which are more common than the
high amplitude peaks which are thought to form normal galaxies
(Eq.~\ref{sigma_delta}), there is a dramatic increase in the number
density of galaxies with decreasing surface density.  Depending on the
choice of bias, there are $10-100$ times more young low surface
density galaxies than normal galaxies, assuming that all galaxies form
with $\lambda=.075$.

If instead we had assumed that all galaxies form with a much smaller
value of $\lambda$, say $\lambda=0.03$, then the total number density
of galaxies formed by $z=0$ would have been a factor of 10 below the
observed number density of normal galaxies.  Because galaxies with
low angular momenta have very large collapse factors, they must have
collapsed from very large shells ($r_0>>r_*$).  Such shells are
rare and have long collapse times, causing the reduced number density
seen if Figures \ref{bias1}\&\ref{bias2.5}.

This points to a hidden assumption in the calculation.  By using the
Press-Schechter formalism, we are expressly counting galaxies at the
time when their halo collapses and virializes.  Thus, there is an
implicit assumption that the collapse of the baryons occurs
simultaneously with the collapse of the halo.  For the collapse of the
baryons and the dark matter to be asynchronous, the baryons must
collapse significantly faster than the free-fall time from the radius
of maximum expansion ($=2\Lambda r_*$).  Such large amounts of
dissipation could only occur after gravity had organized the baryons
into a dense enough system for radiative shocks and star formation to
take place, in other words, after a gravitational collapse time; this
suggests that asynchronous collapse of the baryons and halo is not a
problem for the Press-Schechter system of accounting.  On the other
hand, a fragmentary collapse, where baryons clump during infall, could
speed both dissipation and angular momentum loss; this may be the formation
pathway for spheroids, perhaps leaving globular clusters as debris.

While from the first two columns of Figures 1 \& 2 it is clear that
galaxies of very low surface densities form in large numbers, it is
ambiguous whether or not the mass (or luminosity) of the galaxies has
any leverage when weighed against the mass in normal galaxies.  The
baryonic mass density, $\rho({\bar\Sigma}|z)$, may be calculated by
multiplying $n({\bar\Sigma}|z)$ by the baryonic mass of the galaxies
formed with surface density ${\bar\Sigma}$ (see
equation~\ref{sigma_m}).  Integrating $\rho({\bar\Sigma}|z)$ to get
the cumulative distribution of baryonic density,
$\Gamma({\bar\Sigma}|z)$:

\begin{eqnarray}
\Gamma({\bar\Sigma}|z) &\equiv& \int^{\bar\Sigma}_0 \rho({\bar\Sigma}^\prime|z)
/ F\rho_c \, d{\bar \Sigma}^\prime \\
		     &=& \int^{\bar\Sigma}_0 n({\bar\Sigma}^\prime|z) *
M({\bar\Sigma}^\prime) / F\rho_c \, d{\bar \Sigma}^\prime.
\end{eqnarray}

The resulting distributions for $\rho({\bar\Sigma}|z)$ and
$\Gamma({\bar\Sigma}|z)$ are plotted in the last two columns of
Figures 1 \& 2.  Note that the cumulative density
$\Gamma({\bar\Sigma}|z)$ does not always integrate to 1 because
$\rho({\bar\Sigma}|z)$ does not include mass from shells that collapse
to radii outside of $r_*$ or that have not yet collapsed to $r_*$.
For small $\Lambda$ (short collapse times), the cumulative density is
much larger, as there has been ample time for most overdensities to
collapse.  The small normalization problem for low $\Lambda$ models
reflects the inaccuracy of the approximation to the CDM power spectrum
at small $r_0$.

One may read off the density contributed by galaxies in a
particular range of surface density; the change in
$\Gamma({\bar\Sigma}=\infty|z)$ between any $z_1$ and $z_2$ is the mass
density in galaxies between the corresponding minimum surface
densities at those redshifts, ${\bar\Sigma}_0(z_1)$ and ${\bar\Sigma}_0(z_2)$.
Therefore, we may read off the mass density in low surface density
galaxies as $P(\infty|z=0)-P(\infty|z=2)$, and compare it to the mass
density in normal galaxies $P(\infty|z=2)$.  The mass density of
galaxies with ``sub-normal'' surface density is comparable to or
significantly greater than the mass density in normal galaxies.  In
particular, for models with high bias the fraction of the total mass
density that is tied up in LSB's is dramatically large.  This suggests
that there can easily be as much mass tied up in a population of low
surface brightness galaxies as there are in normal galaxies!  Thus,
LSBs have all the properties that a theorist could hope for: they are
almost entirely unconstrained by observations although they are known
to exist, {\it and} they may contain a large fraction of mass of the
universe.

To develop an appreciation for the impressively low surface brightness
implied by Figures 1 \& 2, consider the spread in the surface
brightnesses of recently formed galaxies.
If $350\surfden$ is the baryonic surface density for normal spirals, then
Figures 1 \& 2 show that galaxies formed between $z=2$ and the
present may have surface densities that are several hundred times
smaller than normal galaxies.  Assuming a linear relationship between
surface density and surface brightness, this corresponds to a
$6\surfb$ range in surface brightness, implying a cosmologically
significant population of galaxies with central surface brightnesses
in $V$ of $27\surfb$!  This range in surface brightness is being
probed in a survey which should soon yield interesting measures of the
density of LSB galaxies (Dalcanton 1995).  Unfortunately, at much
lower surface brightnesses the signal from LSBs may easily be swamped
by fluctuations from very distant cosmologically dimmed clusters of
galaxies (Dalcanton 1995); this will strongly limit the ability of
observations to reveal galaxies with surface brightnesses much fainter
than $V=27\surfb$.

\subsection{On the Effects of Pressure}			\label{jeans}

In the preceeding calculation of the number density of
galaxies as a function of surface density and formation time,
we made an implicit assumption that the baryons will
collapse along with the dark matter halos.  However, unlike
the dark matter (presumably), a baryonic
gas experiences pressure forces in addition to gravitational
forces.  For small masses, internal pressure may support the
gas against collapse, in spite of the inward gravitational
pull of the collapsed dark matter halo (Jeans 1929).

The question arises, then, of whether or not low-surface
brightness galaxies (which we postulate to have low
mass) are capable of collapsing to form stars at
all.  This question may be addressed by considering the
detailed interaction between pressure, gravitational
collapse, and the numbers of collapsed objects,
effectively extending
the Press-Schechter formalism to include the effects
of pressure (Babul 1987, Shapiro et al 1994).  However,
this level of complication may be avoided by rephrasing
the above question to ask: what is the range of surface
densities for which the gas is bound to the dark matter
halo?  The gas will be bound if the gravitational
binding energy is greater than its thermal energy.  Assuming
that the gas is fully ionized, it will be gravitationally
bound to the dark halo if

\begin{equation}					\label{thermgrav}
\frac{m_p V_c^2}{2} > k T,
\end{equation}

\ni where $V_c$ is the circular velocity of the dark matter
halo, $m_p$ is the proton mass, and $T$ is the temperature
of the gas.  The circular velocity is independent
of radius or mass scale if the dark matter collapses to
form an isothermal sphere, and can be expressed as

\begin{equation}				\label{Vc}
V_c^2 = \frac{(18\pi^2)^{1/3}}{2}\,(1+z_c)\,H_0^2\,r_0^2
\end{equation}

\ni (Narayan \& White 1988).  Using Eqns.~\ref{r0_sigma}\&\ref{sigma_z}
to eliminate $z_c$ and $r_0$ in favor of ${\bar \Sigma}$ and $r_*$,
we can express the condition for collapse as a baryonic surface density
threshold:

\begin{equation}					\label{sigthresh}
{\bar \Sigma} > \frac{2 k T F \Lambda}{\pi m_p r_* G},
\end{equation}

\ni where $G$ is the gravitational constant.  Note that the
resulting threshold surface density has no dependence on $z_c$ or
$r_0$.  This reflects that the gravitational potential well of a dark
matter halo depends only on its internal size scale and mass,
which are fully determined by ${\bar \Sigma}$, $\Lambda$, and $r_*$.

The temperature of the gas in the final halo depends on its ability to
cool during collapse.  If the cooling time of the gas is shorter than
the dynamical time of the system, then the gas cools to roughly
$T\approx10^4\,{\rm K}$ at which point cooling becomes highly
inefficient.  Using Figure 3 of Blumenthal et al (1984), we can
examine where forming LSBs are likely to lie with regard to the
``cooling curve'' -- the locus on the temperature-density plane that
delineates the region of parameter space where cooling is rapid
compared to the dynamical time (Rees \& Ostriker 1978).  Blumenthal et
al plot the equilibrium positions of collapsed, non-dissipative
structures (i.e. dark matter halos) on this plane, as a function of
the amplitude of the initial overdensity ($0.5\sigma$ - $3\sigma$).
Even for the small amplitude overdensities which are the likely LSB
precursors, the gas in halos with masses between roughly a few times
$10^8\msun$ and $10^{12}\msun$ is capable of rapid cooling to
$T\approx10^4\,{\rm K}$.  Associating normal galaxies with halo masses
of $>10^{10}\msun$, and taking a naive proportionality between mass
and surface brightness (Eqn.~\ref{sigma_m}), there is a factor of
roughly 100 in surface density over which cooling is efficient.  For a
direct proportionality between surface density and surface brightness,
this corresponds to a disk central surface brightness of 5 magnitudes
below the Freeman value, or $\mu_0(V) = 26.5\surfb$.  Therefore, over
an enormous range of surface brightnesses, we may assume that the
baryonic gas in the halo cools to $T\approx10^4\,{\rm K}$.

Subsituting this temperature into Eqn.~\ref{sigthresh}, we find
a baryonic surface density threshold of

\begin{equation}					\label{sigthreshnum}
{\bar \Sigma} > 0.61 \msun/\pc^2 \, \left(\frac{\Lambda}{10}\right)
                         \, \left(\frac{F}{0.05}\right)
                         \, \left(\frac{10\kpc}{r_*}\right)
                         \, \left(\frac{T}{10^4\,{\rm K}}\right).
\end{equation}

\ni Comparing to Figures 1\&2, this threshold lies below the lowest
surface density expected to be collapsing today and therefore, none of
the baryons in these galaxies should be pressure supported against
collapse.  There is a slight problem with self-consistency, however,
in our assumption of a $10^4\,{\rm K}$ gas temperature throughout this
range of surface density.  The assumption of efficient cooling breaks
down for surface densities that are roughly a factor of 100 below the
surface density of normal galaxies.  In \S\ref{surfbriden} we
associated normal galaxies with surface densities of a few times
$10^2\msun/\pc^2$, and thus the regime of efficient cooling breaks
down at a few $\msun/\pc^2$.  Thus, the likely cut-off in the
distribution of galaxy surface densities is somewhat higher than given
in Eqn.~\ref{sigthreshnum}, and thus about a factor of three
higher than the smallest surface density found in Figures 1\&2.  Note,
however, that this slight change in cut-off hardly changes the
conclusions of the previous section; there are still enormous numbers
of low surface density galaxies for every normal galaxy, as well as
substantial masses tied up in these galaxies.

\section{The Role of Dwarf Galaxies}			\label{dwarfs}

The discussion above has focussed on large galaxies.  The choice
of $r_*\approx10\kpc$ has led us to consider only galaxies with scales
that are typical of normal spiral and elliptical galaxies.  However,
there are also many smaller dwarf galaxies of both high and low
surface brightnesses which exist at the present day.  We now turn
our discussion to these galaxies.  Although galaxies exist with
a continuum of sizes, we will use the term ``dwarf galaxy'' to
refer to galaxies with sizes typical of the sub-$0.1L_*$ galaxies
in the local group, effectively the Large Magellenic Cloud and
smaller.

The formation and subsequent evolution of dwarf galaxies is
necessarily more complicated than for large galaxies.  First,
star-formation in dwarf galaxies is more subject to interruptions due
to supernovae, ionization, or ram-pressure stripping than their more
massive counterparts (see Dekel \& Silk 1986, DeYoung \& Heckman 1994,
Efstathiou 1992).  Secondly, there are other mechanisms besides
gravitational collapse which are capable of producing dwarf galaxies,
for example, clumping in tidal debris during mergers (Barnes \& Hernquist 1992,
Mirabel, Dottori, \& Lutz 1992, Mirabel, Lutz, \& Maza 1991,
Elmegreen, Kaufman, Thomasson 1993), and compression of the intergalactic
medium (Silk, Wyse, \& Shields 1987). Finally, if structure grows
hierarchically, then many of the dwarfs that exist at early times
merge together and are assimilated into larger galaxies by the
present time.  There is a wealth of papers that treat the formation
and evolution of dwarf galaxies in more detail than we are capable
of doing justice to here.  Recognizing our limitation in treating
only the gravitational physics of dwarf galaxies, we restrict ourselves
to a general discussion of how dwarf galaxies fit into our scheme
of early-forming HSB's and late-forming LSB's.

Because dwarf galaxies in principle collapse from smaller initial radii
$r_0$ than do normal galaxies, they in general collapse at earlier
times.  Similarly to large galaxies, dwarfs that collapse from
high-amplitude peaks will collapse at earlier times and to higher
surface densities than those dwarfs which collapse from low-amplitude
peaks.  However, while we were somewhat justified in neglecting
merging of large galaxies, by no means are we
justified in doing so for dwarfs.  Some large fraction of the dwarfs
that exist at large redshifts will merge together and have disappeared
by the present day.  The high surface density dwarfs have the earliest
formation times, the largest amplitude correlation function and thus
the largest probability of being absorbed into the large galaxy
population.  Late-forming, weakly correlated, low surface density
dwarfs have the smallest chance of being absorbed.  This will tend to
deplete the distribution of dwarf surface brightnesses at the high
surface brightness end.  Dense dwarfs are less easily disrupted than
tenuous low surface density dwarfs, however, which may help counteract
the depletion of high surface density dwarfs.  Obviously even the most
simple treatment dwarf galaxies is complicated, and any quantitative
discussion lies far outside of the scope of this paper.

At first blush it appears that our scenario as presented is overruled
by the presence of some old dwarf galaxies with low surface
brightnesses in the Local Group (see Ferguson \& Binggeli (1994)
and references therein).  However, by assuming the
Press-Schechter distribution function, we are implicitly concerning
ourselves with the average number density of collapsed galaxies,
independent of environment.  A much more detailed treatment of the
conditional, environmentally dependent number density by Bower (1991)
shows that galaxies which exist in groups today collapse earlier than
do galaxies in less dense environments.  While our treatment
effectively assumes perfectly synchronized formation times for
galaxies of a particular surface brightness, the Bower formalism
shows how this co-evality breaks down when one considers the range of
environments in which galaxies form.  Therefore, we are not bothered
by the presence of genuinely old low surface brightness dwarfs in the
Local Group.
Furthermore, even in our simple picture, dwarf galaxies which
are three orders of magnitude lower surface brightness than their
early forming high surface brightness counterparts can still form
at $z=2$, a high enough redshift for their stellar populations to
label them as ``old'' systems.

\section{Relevance to Other Astrophysical Issues}	\label{otherastro}

We have postulated that hierarchical structure formation
models naturally lead to a large population of low surface
brightness galaxies.  Such a pervasive population of LSB's
must manifest itself in many astrophysical contexts; we consider
several of these here.

\subsection{HI Surveys}				\label{HI}

It has often been considered a failing of hierarchical structure
formation scenarios that deep HI surveys have failed to uncover a
significant population of dwarf galaxies.  The few uncataloged dwarfs
that are discovered are preferentially found near bright galaxies (see
van Gorkum 1993 for a recent review).  There does not seem to be a
large highly uncorrelated population of gas rich dwarfs.

However, in light of recent work showing a sharp cutoff in HI disks at
column densities of $10^{19}\cm^{-1}$ (van Gorkum et al. 1993,
Corbelli, Schneider, \& Salpeter 1989), the paucity of HI dwarfs is
not surprising.  Recent work by Maloney (1993) and Corbelli \& Salpeter
(1993) convincingly demonstrates that ionization of the HI by the UV
background accounts for the sharp cutoff in HI disks.  As we have shown
that low-mass galaxies tend to have low surface densities, these
galaxies will be prone to having their hydrogen ionized, reducing
their detectable HI masses well below their total hydrogen masses.
Taking Maloney's (1993) scaling for the critical column density
$N_{cr}\propto V_c^{0.5} {\bar\Sigma}_H^{-0.6}$ , (where $V_c\propto
{\bar\Sigma}_0^{1/3}$ is the halo circular velocity, ${\bar\Sigma}_H \propto
{\bar\Sigma}_0$ is the halo surface density, and ${\bar\Sigma}_0$ is the
central
HI surface density), a galaxy's hydrogen will be completely ionized for
${\bar\Sigma}_0 \lta 4 \times 10^{19} \cm^{-2}$ .  Field spirals have
central HI column densities of roughly $10^{20} - 10^{21} \cm^{-2}$
(Cayatte et al. 1993), so galaxies that have surface densities roughly
one-tenth below normal are likely to be highly ionized and thus have
extremely low detectable HI masses.  The observable HI mass can be
expressed as:

\begin{equation}
M_{HI} = 2\pi {\bar\Sigma}_{0} \alpha^{2} \left[1 -
0.01 \left( { {10^{21}} \over {{\bar\Sigma}_{0}}} \right)^{1.43}
\left(5.605 + 1.43 \ln{\left({{\bar\Sigma}_{0}}
\over {10^{21}} \right)}\right) \right]
\end{equation}

\ni A galaxy with an exponential scale length of $\alpha = 5 \kpc$ and a
central HI surface density of $10^{21} \cm^{-2}$ has an HI mass of
$10^{9} \msun$ .  Another galaxy with a central surface density of
$10^{20} \cm^{-2}$ has an HI mass of $4 \times 10^{7} \msun$, two
and a half times lower than expected based on its total hydrogen
surface density  and, more importantly, well below most limits of HI
surveys.  If the central HI surface density were to be dropped by
another factor of two, the HI mass would fall by a factor of sixteen
($\sim 3 \times 10^6 \msun$).

Low surface density galaxies are therefore likely to suffer from strong biases
against their detection in HI surveys, not just in optical surveys.  A
large population of LSB's could easily have been overlooked by existing
surveys.  The dwarfs that have been detected in HI surveys must have
higher surface densities in general, and thus are more likely to have
collapsed earlier from larger overdensities.  This would explain why these
dwarfs are found to trace the bright galaxy population.

The absence of neutral gas in low surface density galaxies is likely
to suppress the star formation in these galaxies, or at least to drive
it through different channels than in the Milky Way.  It is possible
that star formation in low surface density galaxies takes place within
small self-shielding clumps, embedded in the diffuse ionized
background; the calculation of the critical hydrogen surface density
(Maloney 1993) does not take strong clumping into account.

Finally, we note two mitigating circumstances that may improve the prospects
for detecting HI in LSBs.
First, if the star formation efficiency is depressed in LSB
galaxies, they may have a higher gas fraction than normal galaxies, and
thus larger column densities than one might expect from their surface
brightness alone.  Second, because LSBs are more likely to be young
systems, they have had less time to convert gas into stars, also
increasing their gas fraction.  With a higher gas fraction, LSBs would
have larger hydrogen masses and lower ionization fractions than
one would derive from naively scaling the properties of spiral
disks to LSB disks.

\subsection{Lyman-$\alpha$ Absorbers}

Given that LSBs are gaseous and numerous, they must contribute to the
Lyman-$\alpha$ forest.  They are similar to minihalos in that they are
gravitationally-confined systems that have collapsed from small
overdensities.  However they differ from minihalos in many important
respects.  LSBs are disklike systems supported by rotation, whereas
minihalos are assumed to be spherical and supported against
gravitational collapse only by their thermal pressure.  Press \&
Rybicki (1993) have shown that the observed line widths of the
Lyman-$\alpha$ forest are too large to be explained in a model where
the minihalos are in thermal equilibrium with the UV background.  They
argue that the best explanation for the large line widths is not
thermal processes but bulk motion of the gas within the cloud.  While
other workers have considered collapse of a spherical cloud as a
source of these additional velocities, we believe that internal
rotation is an equally plausible, well-motivated explanation.  We will
be addressing this idea in a later paper (Spergel \& Dalcanton 1995).

\subsection{Faint Blue Galaxies}

Because of their blue colors, weak correlations, and underrepresentation
in catalogs of nearby catalogs, LSBs are a natural candidate for
the ``excess'' faint blue galaxies seen in deep galaxy surveys
(McGaugh 1994 -- see references within
for exhaustive listings of the body of work on the faint
blue galaxy excess).  The late formation
times suggested by observations and by the arguments in this paper
are not in conflict with the possibility that LSBs make up the excess
galaxies seen at moderate redshifts; while LSBs as they have been defined
in this paper are formed later than spirals, there are sufficient
numbers of them at the moderate redshifts ($z\lta0.7$) where the
brighter of the faint blue galaxies are found.

We note that a constraint exists on the amount of ``missing'' LSBs
needed to resolve the discrepancy in the number counts.  Dalcanton
(1993) found that the rest frame $B$ luminosity density at
$z\approx0.4$ is $3-5$ times higher than the local luminosity density
as measured in surveys of nearby galaxies.  If this discrepancy is due
entirely to underestimating the contribution of LSBs to the local
luminosity density, then there must be more than a factor of two
greater luminosity density in uncataloged LSBs than in cataloged high
surface brightness galaxies.  The results of the over-simplified model
presented in \S\ref{surfbriden} do not rule out this possibility.

\subsection{Tully-Fisher}

The Tully-Fisher relationship will change as galaxy samples are
extended to lower surface brightness.  The Tully-Fisher relation is an
artifact of the limited range of surface brightnesses sampled in large
galaxy catalogs.  ${\bar\Sigma} \propto M/R^2$ and $V_c \propto
M^{1/2}/R^{1/2}$
imply: ${\bar\Sigma} M \sim V^4$.  This expression reduces to the
Tully-Fisher relation,

\begin{equation}				\label{tfeqn}
L \propto \frac{V^4}{(M/L) \, {\bar\Sigma}}
\end{equation}

\ni where

\begin{eqnarray}
M/L &=& \frac{M_{tot}(r<r_*)}{M_{baryons}} \times \frac{M_{baryons}}{M_{*}}
\times
       					\frac{M_{*}}{L}		\\
\end{eqnarray}

\ni and where $M_{baryons}$ is the total baryonic mass of the galaxy,
$M_*$ is the mass converted into stars, and $M_{total}$ is the total
mass within the region of HI emission (i.e. within the maximum
extent of the baryons).  Taking the definition of the collapse factor
in Eq.~\ref{collapse},

\begin{eqnarray}				\label{mtot_mb}
\frac{M_{tot}(r<r_*)}{M_{baryons}} &=& \frac{1}{F\Lambda}
\end{eqnarray}

\ni for an isothermal dark matter halo.  Referring to Eq.~\ref{sigma_z},
note that Eq.~\ref{mtot_mb} slightly reduces the dependence of
Eq.~\ref{tfeqn} on ${\bar\Sigma}$.

The equations above imply that low
surface brightness galaxies will in general follow the slope of the
Tully-Fisher relationship, but may be offset from the track followed
by normal spirals.  The lower surface density suggests that LSB's will
tend to be overluminous when compared to normal spirals with the same
size and circular velocity.  However, a concomitant reduction in the
star formation efficiency, as might be expected with decreasing
surface density, would pull the luminosity of LSB's downwards.
Variations in the baryonic collapse factor will also contribute to
variations in $M/L$; galaxies which have undergone very little
collapse will have a smaller baryonic mass fraction (i.e.\ large
$M_{tot}/M_{baryons}$), and thus larger mass-to-light ratios.  The
combination of lower surface density and reduced star formation
efficiency may conspire to leave LSB's on the same relationship
followed by normal spirals.  This would be a most fortunate
coincidence for extending Tully-Fisher to larger distances.
Recent work by Sprayberry et al (1995a) is beginning to shed light
on these issues.

We expect much more scatter in the Tully-Fisher relation for LSB
galaxies.  Because of their reduced halo masses and low surface
densities, the disks of LSBs will be much puffier than the disks of
normal spirals, which will lead to much greater uncertainty in the
inclination correction, especially for smaller galaxies.  The
characteristic disk scale height, $Z_0$, is proportional to
$\sigma_{vz} V_c / 4G{\bar\Sigma} $ where $\sigma_{vz}$ is the
vertical velocity dispersion, $V_c$ is the circular velocity at
infinity, and ${\bar\Sigma}$ is the surface density.  Because $V_c^2
\propto M$ and $ M \propto {\bar\Sigma} $,

\begin{equation}
z_0 \sim \frac{\sigma_{vz}}{{\bar\Sigma}^{1/2}},
\end{equation}

\ni suggesting that LSBs will have larger scale heights than normal galaxies
with the same radial extent (modulo the uncertainty of dependence of
vertical temperature on surface density).  The resulting
uncertainty in the inclination correction will be largest for dwarf
LSBs which, while flattened, may hardly look disk-like at all.

\section{Summary}

Hierarchical models of galaxy formation provide a qualitatively and
quantitatively reasonable explanation for many of the global
properties of $L_*$ galaxies.  Analytical models and N-body
simulations can correctly predict the number and luminosity of bright
galaxies as well as their kinematics.  However, these models have
consistently had difficulty in matching the observed faint end of the
luminosity function.  We have shown in this paper that low mass
galaxies tend to form naturally with low surface densities.  Low
surface density galaxies can be assumed to have low surface
brightnesses as well, unless they also have particularly high star
formation efficiencies -- a situation we consider unlikely.  As such,
the faint galaxies predicted by hierarchical clustering models are
likely to have very low surface brightnesses, possibly hundreds of
times fainter than the surface brightnesses of normal spiral disks.
The observed distribution of central surface brightnesses for spirals
suggests that there are selection effects which have lead to dramatic
underestimates of the numbers low surface brightness galaxies;
possibly only one tenth of the galaxies with surface brightnesses of
half of ``normal'' have been cataloged (Disney 1976, Allen \& Shu
1979).  We have argued that the undercounted galaxies are
preferentially low mass, and therefore the failure of models to match
observations of the faint end of the luminosity function should not be
surprising; low mass galaxies are hidden not only by their low total
luminosity, but by their low surface brightnesses as well.
Furthermore, these galaxies should be as difficult to detect in the
radio as they are in the optical.  The gas in galaxies with extremely low
surface densities may be easily ionized by the UV background, effectively
making these galaxies invisible in HI.

In the scenario which we have developed, LSB's collapse slowly and at
late times from small initial overdensities (Mo et al. 1994).  We
argue that galaxies formed from small overdensities are likely to have
larger spin parameters $\lambda$ and thus smaller collapse factors,
leading to lower baryonic surface densities.  Because of the longer
formation timescales, increased spin parameters, decreased
correlations, and decreasing surface brightness, LSBs may be
interpreted as a natural extension of the Hubble sequence, from Sc's
to Sd's and beyond.  However, one would expect LSBs
to appear scruffier than the galaxies which define the bulk of the Hubble
sequence.  The decreased surface density implies that LSBs will tend
have thicker disks than normal galaxies, and thus a decreased baryonic
density within the galaxy.  This in turn reduces the efficiency of
star formation as well as the likelihood that the galaxy becomes
unstable to spiral structure.  The low surface density and low overall
mass also increase the likelihood that feedback from star formation
through supernova-driven winds can affect the apparent morphology of
LSB galaxies.  All of these processes may lead to the diffuse, chaotic
galaxies which are becoming associated with the extremes of galaxy
surface brightness (Dalcanton 1995, Schombert et al 1992).

Although LSBs are hardly impressive members of the galactic community
when viewed as individuals, their cumulative properties are
impressive.  Both the number density and mass density of the LSB
population are substantial, with LSBs possibly contributing as much
(or more) mass to the universe as normal galaxies.  (Although we have
derived this result specifically assuming a CDM spectrum of initial
fluctuations and $\Omega=1$, our qualitative results should be little
changed by assuming a different cosmological model, as long as
structure formation proceeds hierarchically.)  Given this possibility,
an accurate measure of the number density and mass density of the LSB
population becomes an important goal for the coming years.  While LSBs
may be overlooked as individuals, they are potentially too important
to be overlooked as a class.

\bigskip
\bigskip
\centerline{\bf Acknowledgements}
\medskip

We thank Daniel Eisenstein and Avi Loeb for providing the data used to
generate the distribution of $\lambda$ shown in Eisenstein \& Loeb
(1994) and for discussions on the origins of spin angular momentum, and
Arif Babul for his insights on the effects of pressure on galaxy formation.
Jim Gunn is also warmly thanked for discussions which, although
typically brief, are always enlightening and enjoyable.
JJD and DNS
were supported by NASA and by a Hubble Fellowship.
Ron Kim and Teresa Shaw are
warmly thanked for assistance with typing during the Month of Pain, as
is Dan Wallach for the loan of the keyboard adapter.

\vfill

\section{References}

\hi{Allen, R. J., \& Shu, F. H. 1979, \apj 227, 67.}

\hi{Babul, A. 1989, Ph.D. thesis, Princeton University.}

\hi{Babul, A., \& Rees M. J. 1992, \mn 255, 346.}

\hi{Barnes, J. E., \& Hernquist, L. 1992, \nat 360, 715.}

\hi{Barnes, J. E., \& Efstathiou, G. 1987, \apj 319, 575.}

\hi{Bernstein, G. M., Tyson, J. A., Brown, W. R., \& Jarvis, J. F. 1993,
\apj in press.}

\hi{Bond, J. R., Cole, S., Efstathiour, G. \& Kaiser, N. 1991, \apj 379, 440.}

\hi{Bothun, G. D., Beers, T. C., Mould, J. R., \& Huchra, J. P. 1986,
\apj 308, 510.}

\hi{Bothun, G. D., Impey, C. D., \& Malin, D. F. 1991, \apj 376, 404.}

\hi{Bothun, G. D., Impey, C. D., Malin, D. F., \& Mould, J. 1987, \aj 94, 23.}

\hi{Bothun, G. D., Schombert, J. M., Impey, C. D., Sprayberry, D.,
McGaugh, S. S. 1993, \aj 106, 530.}

\hi{Bower, R. J. 1991, \mn 248, 332.}

\hi{Dalcanton, J. J. 1993, \apjl 415, L87.}

\hi{Dalcanton, J. J. 1994, submitted to {\it Ap. J.}}

\hi{Dalcanton, J. J. 1995, in preparation.}

\hi{Davis, M., Efstathiou, G., Frenk, C., \& White, S. 1985, \apj 292, 371.}

\hi{Davies, J. I., Phillipps, S., \& Disney, M. J. 1990, \mn 244, 385.}

\hi{Dekel, A., \& Silk, J. 1986, \apj 303, 39.}

\hi{DeYoung, D., \& Heckman, T. 1994, \apjl 431, 598.}

\hi{Disney, M. 1976, \nat 263, 573.}

\hi{Eisenstein, D. J., \& Loeb, A. 1994, \apj submitted.}

\hi{Efstathiou, G. 1992, \mn 256, 43p.}

\hi{Efstathiou, G. \& Silk, J. 1983, \fcp 9, 1.}

\hi{Efstathiou, G., Bernstein, G., Katz, N., Tyson, J. A., \& Guhathakurta, P.
1991, \apjl {\bf380}, L47.}

\hi{Elmegreen, B. G., Kaufman, M., Thomasson, M. 1993, \apj 412, 90.}

\hi{Evrard, A. E., Summers, F. J., \& Davis, M. 1994, \apj 422, 11.}

\hi{Faber, S. M. 1982, in {\it Astrophysical Cosmology}, ed. H. A. Bruck,
G. V. Coyne, \& M. S. Longair, (Vatican: Pontificia Acadamia Scientiarum),
191.}

\hi{Faber, S. M., \& Gallagher, J. S. 1979, \araa 17, 135.}

\hi{Ferguson, H. C., \& Binggeli, B. 1994, {\it Astronomy \&
Astrophysics Review}, in press.}

\hi{Ferguson, H. C., \& McGaugh S. S. 1995, \apj in press.}

\hi{Freeman, K. C. 1970, \apj 160, 811.}

\hi{Gallagher, J. S.,  \& Hunter, D. A. 1984, \araa 22, 37.}

\hi{Gunn, J. E. 1982, in {\it Astrophysical Cosmology}, ed. H. A. Bruck,
G. V. Coyne, \& M. S. Longair, (Vatican: Pontificia Acadamia Scientiarum),
191.}

\hi{Hoffmann, Y., Silk, J., Wyse, R. F. G. 1992, \apjl 388, L13.}

\hi{Impey, C. D., Bothun, G. D., \& Malin, D. F. 1988, \apj 330, 634.}

\hi{Impey, C. D., Sprayberry, D., Irwin, M., \& Bothun, G. D. 1993, preprint.}

\hi{Irwin, M. J., Davies J. I., Disney, M. J., \& Phillips, S. 1990, \mn
245, 289.}

\hi{Jeans, J. H. 1929, {\it Astronomy and Cosmogony,} 2nd ed., (Cambridge,
England: Cambridge University Press).}

\hi{Kaiser, N. 1984, \apjl 284, L9.}

\hi{Kennicutt, R. C. 1989, \apj 344, 685.}

\hi{Koo, D. C., \& Szalay, A. 1984, \aj {\bf282}, 390.}

\hi{Knezek, P. 1993, Ph.D. Thesis, University of Massachusetts.}

\hi{Lifshitz, E.M. 1946, {\it Zh. Eksp. Teoret. Fiz.}, 16, 587.}

\hi{Lo, K. Y., Sargent, W. L. W., \& Young, K. 1993, \aj 106, 507.}

\hi{Marzke, R. O., Geller, M. J., Huchra, J. P., \& Corwin, H. G.
1994, \aj in press.}

\hi{McGaugh, S. S., \& Bothun, G. D. 1994, \aj 107, 530.}

\hi{McGaugh, S. S. 1994, \apj 426, 135.}

\hi{McGaugh, S. S. 1994, \nat 367, 538.}

\hi{McGaugh, S. S., Schombert, J. M., \& Bothun, G. D. 1995, \aj in press.}

\hi{Mirabel, I. F., Dottori, H., \& Lutz, D. 1992, \aa 256, L19.}

\hi{Mirabel, I. F., Lutz, D., \& Maza, J. 1991, \aa 243, 367.}

\hi{Mo, H. J., McGaugh, S. S., \& Bothun, G. D. 1994, \mn 267, 129.}

\hi{Narayan, R., \& White, S. D. M. 1988, \mn 231, 97p.}

\hi{Nilson, P. 1973, Uppsala General Catalog of Galaxies, {\it Uppsala
Astr. Obs. Ann.}, Vol.6.}

\hi{Peacock, J. \& Heavens A. 1989, \mn 243, 133.}

\hi{Peebles, P. J. E. 1969, \apj 155, 393.}

\hi{Phillipps, S., Davies, J. I., \& Disney, M. J. 1990, \mn 242, 235.}

\hi{Press, W. H., \& Schechter, P. L. 1974, \apj 330, 579.}

\hi{Pritchett C. J., \& Infante L. 1992, \apjl {\bf399}, L35.}

\hi{Rees, M. J., \& Ostriker, J. P. 1978, \mn 179, 541.}

\hi{Roberts, M. S. 1969, \aj 74, 859.}

\hi{Schombert, J. M., \& Bothun, G. D. 1988, \aj 95, 1389.}

\hi{Schombert, J. M., Bothun, G. D., Schneider, S. E., \& McGaugh, S. S. 1992,
\aj 103, 1107}

\hi{Shapiro, P. R., Giroux, M. L., \& Babul, A. 1994, \apj 427, 25.}

\hi{Shectman, S. A. 1974, \apj 188, 233.}

\hi{Silk, J., Wyse, R. F. G., \& Shields, G. A. 1987, \apjl 322, L59.}

\hi{Sprayberry, D., Bernstein, G. M., Impey, C. D., \& Bothun, G. D. 1995a,
\apj 438, 72.}

\hi{Sprayberry, D., Impey, C. D., Bothun, G. D., Irwin, M. J. 1995b, \aj 109,
558.}

\hi{Steinmetz, M., \& Bartelmann, M. 1994, \mn submitted.}

\hi{Stevenson, P. R. F., Shanks, T., Fong, R., \& MacGillivray, H. T. 1985,
\mn {\bf213}, 953.}

\hi{Thuan, T. X., Alimi, J., Gott, J. R., \& Schneider, S. E. 1991, \apj
370, 25.}

\hi{Thuan, T. X., \& Seitzer, P. O. 1979, \apj 231, 680.}

\hi{Turner, J. A., Phillips, S., Davies, J. I., \& Disney, M. J. 1993, \mn
261, 39.}

\hi{van der Hulst, J. M., Skillman, E. D., Smith, T. R., Bothun, G. D.,
McGaugh, S. S., \& de Blok, W. J. G. 1993, \aj 106, 548.}

\hi{Warren, M. S., Quinn, P. J., Salmon, J. K., \& Zurek, W. H. 1992, \apj
399, 405.}

\hi{White, S. D. M. 1994, MAP preprint.}

\hi{White, S. D. M., David, M., Efstathiou, G., \& Frenk, C. S. 1987, \nat
330, 451.}

\hi{White, S. D. M., \& Frenk, C. S. 1991, \apj 379, 52.}

\hi{White, S. D. M., \& Rees, M. J. 1978, \mn 183, 341.}

\hi{Zwicky, F. 1957, in {\it Morphological Astronomy}, (New York:
Springer-Verlag).}

\vfill
\clearpage

\newcounter{figcnt}

\begin{list}
{Figure \arabic{figcnt}.}  {\usecounter{figcnt}}

\item The number and mass density of galaxies as a function of
surface brightness and redshift.  The lightest line in each frame
corresponds to $z=0$, with each successively darker line corresponding
to $z=1$, $z=2$, etc, up to $z=5$.  Each row of plots corresponds to a
single value of $\lambda$: 0.15 in the top row, 0.075 in the middle,
and 0.03 in the bottom row.  The leftmost column shows the number
density of galaxies with surface brightness ${\bar\Sigma}$ in units of
$\numden$.  The second column is the cumulative distribution of the
number density shown in the first column.  The horizontal dashed lines
are the integrated number density in normal galaxies with $L\L_*>0.1$
for all galaxies (long dash), for spirals only (short dash), and for
spirals Sc and later (dotted line), taken from Marzke \et (1994).  The
third column is the baryonic mass density in galaxies of surface
brightness ${\bar\Sigma}$, given in units of $1/F\rho_c$, the total baryonic
mass density.  The rightmost column is the integrated mass density.
The $1-10\%$ error in normalization in the integrated mass density for
$\lambda=0.15$ reflects the failure of the approximation to the CDM
power spectrum at $r_0<50kpc$.  As discussed in the text, normal disk
galaxies have ${\bar\Sigma}\approx250-400\surfden$.  The value of $r_*$ was
chosen to give the correct number density in galaxies formed before
$z=2$ for $\lambda=0.075$, the spin parameter appropriate for
spiral galaxies.
\label{bias1}

\item Same as Figure~\ref{bias1}, except that $b=2.5$ and $r_*=7.5\kpc$.
\label{bias2.5}

\end{list}

\vfill
\clearpage



\end{document}